\documentstyle[12pt]{article}
\begin{document}
\baselineskip 0.3in

\vspace{1.1in}

\begin{center}
{\LARGE{\bf Investigation of acceptor levels and hole scattering
mechanisms in p-gallium selenide by means of transport measurements
under pressure}}

\vspace{0.5in}
{\large D.Errandonea$^{\dagger}$, J.F.S\'anchez-Royo and A.Segura

\vspace{0.2in}
Institut de Ci\`encia dels Materials, Universitat de Val\`encia\\
Dpto. de F\'{\i}sica Aplicada, Ed. Investigaci\'o, 
Univ. de Val\`encia,\\
C/Dr. Moliner 50, E-46100 Burjassot (Val\`encia), Spain

\vspace{0.4in}
A.Chevy and L.Roa

\vspace{0.2in}
Laboratoire de Physique des Mileux Condens\'es, Univ.
Pierre et\\ 
Marie Curie, 4 place Jussieu, 75252 Paris, Cedex 05, France}
\end{center}

\vspace{.7in}
\begin{abstract}
The effect of pressure on acceptor levels and hole scattering 
mechanisms in $\mbox{p-GaSe}$ is investigated through Hall effect 
and resistivity measurements under quasi-hydrostatic conditions up to 
4 GPa. The pressure dependence of the hole concentration is 
interpreted through a carrier statistics equation with a single
(nitrogen) or double (tin) acceptor whose ionization energies 
decrease under pressure due to the dielectric constant increase.
The pressure effect on the hole mobility is also accounted for 
by considering the pressure dependencies of both the phonon 
frequencies 
and the hole-phonon coupling $\mbox{constants}$ involved in the 
scattering rates.
\end{abstract}

\vspace{0.4in}

\noindent
{\bf Short title:} Transport properties of p-GaSe under pressure.

\vspace{0.3in}
\noindent
$^{\dagger}$Corresponding author, E-mail: daniel@ges1.fisapl.uv.es,
Fax: 34 6 398 31 46.

\pagebreak
\section{INTRODUCTION}
\indent

Gallium selenide (GaSe) belongs to III-VI layered semiconductor
family which is characterized by a strong anisotropy in the chemical 
bonding. Its crystal $\mbox{structure}$ consists of plane hexagonal 
lattices which are associated in pairs and may be stacked in different
ways($\beta,\gamma,\varepsilon$-polytypes)\cite{1}-\cite{2}, being  
the most common one the corresponding to $\varepsilon$-polytype
($D_{3h}^1$ space group). The possibility of growing III-VI 
semiconductors thin films by van der Waals epitaxy\cite{v1}-\cite{v4} 
opens new opportunities to their potential practical applications, 
e.g. electronic 
devices\cite{v3} and solar cells\cite{nc}. In this respect, 
the study of the fundamental electrical transport properties is a 
very important issue. In the case of GaSe, at ambient pressure, these 
$\mbox{properties}$ have been investigated in detail by different 
authors who 
have focused their $\mbox{attention}$ on the role played by impurities 
such as 
Cd\cite{cd}, Cu\cite{cu}, N\cite{jf}, Sn\cite{jf2} and Zn\cite{zn}. 
In contrast, very little is known about the behaviour of transport 
properties under pressure (P). 

Pressure experiments are an efficient tool to study III-VI
semiconductors since they allow the tuning of the degree of 
anisotropy of these materials. $\mbox{Experiental}$ studies of their 
optical\cite{5}-\cite{12} and lattice dynamical\cite{12}-\cite{14} 
properties under $\mbox{pressure}$ have been  subjects of recent 
interest. 
However, systematic research of the electrical transport properties 
under compression has been only made for $\mbox{indium}$ selenide 
(InSe), for 
which, resistivity ($\rho$) and Hall effect (HE) measurements under 
pressure have been reported very recently\cite{stut}.

In this article, we report a systematic study on transport
properties  under high pressure, up to 4 GPa at room temperature (RT), 
of the III-VI layered p-GaSe. In Section II we briefly describe 
the experimental setup. The data obtained on HE and resistivity 
measurements are presented in Section III. Finally, Section IV is 
devoted to the discussion of these results and their implications.  

\section{EXPERIMENT}
\indent

The p-GaSe single crystals used in this study were grown by the
conventional Bridgman technique. Doping by tin (Sn) was performed by
adding the pure element so as to get 0.05, 0.1, 0.2 and 0.5 at $\%$ 
in the stoichiometric melt of GaSe. Nitrogen (N) was introduced as GaN
compound in a quantity to give a concentration of 0.5 at $\%$ of
N in the melt. We should point out that these concentrations are 
different from the effective ones, since during the Bridgman growth 
most of the initial impurity concentration is segregated to the end 
of the ingot\cite{chevy}.

Samples with faces perpendicular to the c-axis were 
prepared from the ingots by cleaving and cutting with a razor blade.
Thickness of the slabs was measured by using the interference fringe
pattern in the infrared transmission spectrum. Typical sample 
dimensions were 70 $\mu$m in thickness and about 4$\times$4 mm$^2$ in 
size. Pressure up to 4 GPa was applied to the samples by using the 
Bridgman cell that has been described in Reference \cite{stut}. In 
this case, we have used tungsten carbide anvils, 27 mm in diameter,  
which were put between the pistons of a 150-ton press.
 Sodium chloride was the pressure transmitting-medium. The 
 pyrophillite 
gaskets were 0.5 mm in thickness, with a hole of 9 mm. The 
Bridgman gasket assembly and geometry are shown in the inset of Figure 
1. Ohmic contacts were made in a van der Pauw geometry\cite{15} by 
soldering silver wires with high-purity indium in gold contact 
pads which were previously vacuum evaporated. HE and resistivity 
measurements were carried out at RT. Direct electric current 
ranging from 50 $\mu$A to 500 $\mu$A was sent through the samples, 
and the resulting voltages were measured by a digital voltmeter. 
The linearity of the ohmic voltages on the injected current was 
checked out at different pressures. A magnetic field of 0.6 Tesla 
was applied parallel to the c-axis.
The magnetic field was generated by means of a copper coil placed
around one of the pistons of the press.

\section{RESULTS}
\indent

Figure 1 shows the pressure dependence of resistivity for N-doped and
$\mbox{Sn-doped}$ GaSe samples. The nominal doping concentration ([N] 
or [Sn]) of each measured sample is also given in this figure. 
Resistivity
appears to decrease with pressure in all the samples here studied. 
This evolution is more pronounced in the samples doped with N, 
in which $\rho$ goes down a factor 
three in the explored range of pressure. This 
factor is nearly constant in all the Sn-doped samples, in which 
resistivity decreases with doping, except 
in the slabs with the highest tin content ([Sn] = $\mbox{0.5 $\%$}$). 

The decrease of the resistivity with
pressure turns out to be due to the increase of both the hole
concentration and mobility, as shown in Figures 2 and 3.
Figure 2 shows the pressure behaviour of the hole concentration ($p$)
for different samples as determined through
\begin{eqnarray}
p=\frac{1}{q R_H} \quad ,
\end{eqnarray}

\noindent
where $q$ is the electron charge and $R_H$ is the Hall coefficient. 
The Hall factor has been assumed to be 1\cite{jf,jf2}. It can be seen 
in Figure 2 that $p$ non-linearly increases under compression, the 
relative variation of $p$ being higher for samples doped with N. In 
samples doped with Sn, at zero pressure $p$ is enlarged with 
increasing [Sn] with the exception of the 0.5$\%$ Sn-doped sample in 
which $p$ is two orders of magnitude smaller. In addition, in this 
sample the relative change of $p$ is the largest among the Sn-doped 
samples. 

As regards the mobility ($\mu$), Figure 3 shows its pressure 
dependence. It can be seen there that the mobility also increases with
pressure and that its value at ambient pressure is very similar to 
that obtained in previous works\cite{jf,jf2}.  

\section{DISCUSSION}

\subsection{\underline{Pressure Dependence of the Hole Concentration}}
\indent

To understand the pressure dependence of $p$ we have considered 
that N-doped and Sn-doped GaSe at RT are extrinsic as   reflected
in the temperature dependence of their transport 
properties\cite{jf,jf2}. Moreover, from the temperature behaviour of 
the hole concentration it was deduced that doping by N introduces only
one $\mbox{acceptor}$ level with ionization 
energy of $E_a$ = 210 meV\cite{jf}. 
In the Sn-doped $\mbox{samples}$ two impurity levels with 
ionization energies of $E_a$ = 155 and 310 meV  
have been observed\cite{jf2}. These levels appear to be connected with 
the presence of a double acceptor-impurity. This acceptor has been 
proposed to be an interstitial Sn atom in the octahedral interlayer 
site associated to a close Ga-vacancy, $\mbox{giving}$ a local 
configuration
similar to that of Sn in the layered compound SnSe$_2$\cite{jf2}. 
At RT the first level ($E_a$ = 155 meV) is not fully ionized, so that 
the 
pressure $\mbox{dependence}$ of $p$ is mainly determined by its 
behaviour under pressure. However,  when $\mbox{increasing}$ the 
doping of Sn 
to 0.5$\%$ a decrease of the hole concentration has been observed 
(see Figure 2) indicating that, at very high Sn concentration, a 
donor $\mbox{configuration}$ of Sn in GaSe (isolated Sn interstitial 
or Sn
substituting to Ga) becomes dominant and overcompensates the
acceptor centers. Then, as in the heaviest doped samples the first 
level (155 meV) is completely compensated, the behaviour of the second 
level (310 meV) would determine the variation of $p$ under 
$\mbox{compression.}$ $\mbox{Therefore,}$ 
we can analyze in both cases the pressure dependence of $p$ 
through a single $\mbox{acceptor-single}$ donor model for 
partially compensated 
p-type semiconductors\cite{17}. Within this model the hole 
concentration is given by:
\begin{eqnarray}
p&=&\frac{N_d}{2} \Bigg\{-1-\frac{N_v}{2N_d} 
\exp\left(\frac{-E_a}{kT}\right) +\nonumber\\
\nonumber\\
&&\left[1+\frac{N_v}{N_d} \exp\left(\frac{-E_a}{kT}\right)
\left(-1+\frac{N_v}{4N_d}\exp\left(\frac{-E_a}{kT}\right)
+\frac{2N_a}{N_d}\right)\right]^{1/2}\Bigg\}
\end{eqnarray}

\vspace{0.1in}

\noindent
where $T$ is the absolute temperature, $k$ is the Boltzmann constant,
$E_a$ refers to the ionization energy of the acceptor level,
$N_d$ and $N_a$  are the donor and acceptor impurity concentration, 
respectively, and $N_v$ is the density of states of the valence band 
which can be written at RT as a function of the effective mass in the 
valence band $m_v^* = ({m^*_{h\perp}}^2 m^*_{h\parallel})^{1/3}$ as:
\begin{eqnarray}
N_v = 2.509  \times  10^{19} \left(\frac{m^*_v}{m_o}\right)^{3/2} 
\mbox{cm}^{-3}\quad,\nonumber
\end{eqnarray}

\noindent
where $m_o, m^*_{h\perp}$, and $m^*_{h\parallel}$ stand for the 
free electron mass, the perpendicular and parallel effective 
hole masses, respectively. Taking $m^*_{h\perp}$ = 0.8 
$m_o$ and $m^*_{h\parallel}$ = 0.2 $m_o$ at zero pressure\cite{34,35}, 
 we obtain $m^*_v$ = 0.5 $m_o$.

In Eq.(2)  one can see that $p$ depends on the pressure 
through $m^*_v$, $E_a$ and the term $N_d/2$. Following the Kane 
model\cite{50}, the variation of the effective mass with pressure is 
considered to be proportional to the change of the direct band-gap
\cite{7}. For the pressure dependence of $N_d$ it can be assumed that 
it is determined by the volume variation\cite{12}.  
Then, taking for $N_a$, $N_d$ and $E_a$ at ambient pressure the values 
deduced from the temperature dependence of the hole 
concentration\cite{jf,jf2}, and using Eq.(2), we can calculate
$E_a$ at each pressure. The result is shown in Figure 4 (points).
There it can be seen that the observed behaviour of $p$ would be a
consequence of a reduction of the ionization energies under pressure.
For the same pressure variation (from 0 to 4 GPa) 
the ionization energy of the N-related acceptor would decrease 
by 20 $\%$, but that of the Sn-related
one, only by 6.4 or 8.1 $\%$. 

The ionization energy of a single hydrogenic impurity level can be
evaluated in an anisotropic crystal through the Gerlach-Pollman 
model\cite{pollman}. By using the $\mbox{effective}$ masses of GaSe 
from References \cite{34,35} 
and the static dielectric constant at ambient conditions 
from Reference \cite{12} one obtains a value of $E_a$ = 72 meV. 
The discrepancy between this value and the experiment can be 
attributed to the differences between the Coulomb potential used in 
the hydrogenic model and the true impurity potential. It has been 
shown that the true impurity potential can be represented by an 
effective potential which includes a central-cell correction to 
the Coulomb potential model\cite{ning}. Then, to modelize the pressure 
dependence of $E_a$ we take the impurity potential to be\cite{ning}:
\begin{eqnarray}
V(r)=-\frac{2 Z R^{*} a^{*}}{r} \Big[1 - \exp(-br) + 
B r \exp(-br)\Big] f(\alpha,\theta)\quad,
\end{eqnarray}

\noindent
where $R^{*}$, $a^{*}$ and $f(\alpha,\theta)$ are the effective 
Rydberg energy, the effective Bohr radius and the anisotropy function 
defined in the Gerlach-Pollman model, respectively, $b$ and $B$ are 
the adjustable potential parameters, and $Z$ is equal to 1(2) for the 
single(double) ionized acceptor-impurity. At small $r$, this potential 
looks like a sphericall well, with a depth $V_0$ given by:
\begin{eqnarray}
V_0 = - q^2 \frac{B+b}{\varepsilon_{0\perp}\varepsilon_{0\parallel}}
\quad,
\end{eqnarray}

\noindent
where $\varepsilon_{0\perp}$ and $\varepsilon_{0\parallel}$ are the 
static dielectric constant in the direction $\mbox{perpendicular}$ 
and parallel 
to the c-axis, respectively. At large $r$ it behaves like a 
screened Coulomb potential. The turning point is given approximately 
by
\begin{eqnarray}
r_z=\frac{B+b}{Bb}\quad .
\end{eqnarray}

\noindent
The proposed impurity potential model and the Coulomb potential are 
shown schematically in the inset of Figure 4. 

The Schr\"odinger equation of the system is solved by the variational 
method by using a trial function for
the electron ground state ($\Psi_0(r)$) as:
\begin{eqnarray}
\Psi_0(r) = \sqrt{\frac{b^3}{\pi}} \exp(-br)\quad.
\end{eqnarray}

\noindent
The variational calculation in which the total energy is minimized 
leads to:
\begin{eqnarray}
E_a=-\left(\frac{5}{9}\right)^2 R^* Z_0^2(\alpha) - 
\frac{16}{27}B R^* a^* Z_0(\alpha) \quad,
\end{eqnarray}

\noindent
where $Z_0(\alpha)$ is the effective charge\cite{pollman}. Then the 
potential parameter $B$ is chosen to fit the ionization energy of 
the single acceptor level at room pressure.  The fitted value of $B$
is shown in Table 1 together with $V_0$ and $r_z$.  $E_a$ depends on 
the pressure through $R^*$ and $a^*$. The first term of Eq.(7)
decreases under pressure due to the reduction of $R^*$ as a 
consequence 
of the increase of $\varepsilon_{0\parallel}$\cite{12,28}. The second 
term also decreases under pressure, but more slowly because 
$R^* a^* \sim (\varepsilon_{0\parallel})^{-1/2}$.

The calculated pressure dependence of $E_a$ is shown in Figure 4 
(solid lines). 
It can be seen there that the model fits quite well the 
experiment. The shift observed in the second Sn-related level may be 
connected with the fact that in the present model we have neglected 
the 
electron-electron interactions and, as a consequence of this, we could
be overestimating the value of $E_a$. To understand why $E_a$ 
suffers a greater decrease in N-doped samples, let us consider the 
parameters given in Table~1. 
It can be seen there that in this case the square well is deeper. This 
is just what one may expect because N is much more electronegative 
than Sn. That is why the central-cell correction is more important in 
N-doped samples than in $\mbox{Sn-doped}$ samples. Because of this 
a deeper acceptor level is obtained. In addition, the second term of 
Eq.(7) would decrease more rapidly leading to the greater change 
observed in $E_a$.  

\subsection{\underline{Pressure Dependence of the Hole Mobility}}
\indent

Ionized impurity scattering and two-phonon scattering mechanisms must 
be considered in order to give a quantitative account of the pressure 
dependence of the hole mobility. The ionized impurity concentration
has been assumed to be that obtained from the temperature dependence 
of
the hole concentration\cite{jf,jf2}. The phonons involved are the 138 
cm$^{-1}$ $A_1^{'2}$ homopolar optical mode and the 153.2 cm$^{-1}$ 
$E^{'3}$ LO polar optical mode\cite{12}. The pressure dependence of 
the Fr\"{o}hlich constant\cite{frolich} ($\alpha$) and the 
hole-homopolar phonon coupling constant\cite{fivaz} ($g^2$) were 
calculated from data found in the literature\cite{12,21,22}. 
As the hole scattering rate for LO phonons is obtained through
an integration over all the possible directions of the phonon
momentum, $\alpha$ was calculated through\cite{jm}:
\begin{eqnarray}
\alpha=\frac{2}{3}\alpha_{\perp}+\frac{1}{3}\alpha_{\parallel}\quad.
\end{eqnarray}

\noindent
We have obtained for both coupling constants a decreasing behaviour 
under $\mbox{pressure}$. 
The reduction of $g^2$ is basically a result of the 
increase of the homopolar phonon frequency 
under pressure\cite{12,11}. For $\alpha$, in spite of the increase 
of $\alpha_{\parallel}$, due to the great increase of 
$\varepsilon_{0\parallel}$ under compression\cite{12,28}, the overall
change is negative because of the decrease of $\alpha_{\perp}$, which  
is more than four times higher than $\alpha_{\parallel}$

By means of an iterative method\cite{17} 
we have calculated the pressure $\mbox{dependence}$ of $\mu$. 
The ionized impurity scattering has been included through the
$\mbox{Brooks-Herring}$ relaxation time\cite{brooks}, in which the 
ionized 
impurity concentrations has been assumed to be that used to calculate
$E_a$ a a function of pressure. 
The results obtained for the samples doped with [Sn] = 0.05 
$\%$ and 0.1 $\%$ and with [N] = $\mbox{0.5 $\%$,}$ 
which are plotted in solid 
lines in Figure 3, agree quite well with the measurements. The 
isolated 
contribution of each scattering mechanism is also represented in 
Figure 3. The $\mbox{homopolar}$ phonon 
scattering (curve 1) and the LO polar phonon 
scattering (curve 2) are the dominant mechanisms over the whole 
pressure range, but the ionized impurity scattering mechanism 
(curve 3) must be taken into account to $\mbox{reproduce}$ 
quantitatively the 
absolute value of $\mu$. The saturation of the increase of $\mu$ above 
2.5 GPa is due to the saturation of the decrease of $\alpha$ as a 
consequence of the compensation between the LO polar phonon frequency 
and 
$\varepsilon_{0\parallel}$ increases. In $\mbox{addition}$, 
the increase of the 
concentration of ionized impurities, determined by the 
increase of $p$ and the contraction of the volume under compression, 
leads to a reduction of the impurity-limited mobility, which 
also collaborates to the saturation of $\mu$.

Finally, we want to point out that for samples doped with [Sn] = 0.2
$\%$ and $\mbox{0.5 $\%$}$ the value of the zero 
pressure mobility cannot be
reproduced with the present model. Samples from the 0.2 $\%$ ingot
exhibit a phonon-controlled $\mbox{mobility}$ 
in a larger temperature range than
those from ingots with lower tin $\mbox{content.}$
We think that this can be related
to the fact that a large proportion of complex impurity 
$\mbox{centers}$ are 
present in these samples, resulting in a $\mbox{reduction}$ of the 
$\mbox{concentration}$ of single ionized impurities\cite{jf2}. 
These complex centers appear to have a lower 
$\mbox{scattering}$ cross section
than ionized impurities, producing that $\mbox{scattering}$ by phonons 
is the
dominant scattering mechanism in spite of the heavy doping 
$\mbox{concentration}$ of the 0.2$\%$ Sn-doped sample. Nevertheless, 
the structure of those centers and its influence on the hole mobility
is not known and can hardly be included in our model. In samples from
the 0.5 $\%$ Sn-doped ingot the $\mbox{compensation}$ 
is very high and impurity
scattering is dominant even at room $\mbox{temperature}$ reducing the 
mobility 
to the low value observed ($\mu$ = 9.5 cm$^2$/Vs), but this value and 
its
temperature dependence could not be accounted for through any known
scattering mechanism.

\section{CONCLUSIONS}
\indent

Transport measurements have been carried out under pressure in 
N-doped and Sn-doped GaSe up to 4 GPa. Within the framework of a 
single acceptor-single donor model, the observed 
increase of $p$ under compression has been interpreted to be due 
to the reduction of $E_a$ with pressure. Modeling the impurity 
potential we have obtained an expression for $E_a$. This allows us 
to relate the decrease under pressure of $E_a$ to the increase of 
$\varepsilon_{0\parallel}$. In addition, we have also discussed the 
different behaviour under pressure of the ionization energies of the
acceptor levels connected to Sn and N as doping impurities in GaSe.
The higher reduction of $E_a$ observed in N-doped samples has been 
explained by means of the deeper $\mbox{central-cell}$ correction 
obtained in this case. The observed increase of $\mu$ has been 
attributed to the decrease of the hole-phonon coupling constants. 

\vspace{0.4in}
\noindent
{\large{\bf ACKNOWLEDGMENTS}}

This work was supported by the Spanish Government CICYT under
Grant No: MAT 95 - 0391.

\pagebreak
\large
\noindent

\begin{table}
\begin{itemize}
\item[]{
\parbox{8.5cm}{\caption {Model parameter $B$, deep of the 
square well $V_0$ and turning point $r_z$.}}

\vspace{0.1in}
\begin{tabular}{cccc}
\hline
\hline
Impurity&$B$&$V_0$&$r_z$\\
\hline
Sn$_1$&0.36&-0.725 eV&4.40 $a^*$\\
Sn$_2$&0.36&-1.120 eV&2.47 $a^*$\\
N&0.53&-1.020 eV&4.25 $a^*$\\
\hline
\hline
\end{tabular}}
\end{itemize}
\end{table}

\vspace{5in}
.

\pagebreak
\noindent
{\large {\bf Figure captions}}

\normalsize

\vspace{0.25in}
\noindent
{\bf Figure 1:} Resistivity as a function of pressure for different 
samples. The inicial doping concentrations in the growth solutions
([N] or [Sn]) are indicated in the figure. The inset shows the
Bridgman gasket assembly.

\vspace{0.25in}
\noindent
{\bf Figure 2:} Hole concentration as a function of pressure
for different samples: $\mbox{($\bullet$ and 
$\vrule height4pt width4pt$)}$ 
[N] = 0.5 $\%$, ($\nabla$) [Sn] = 
0.5 $\%$, ($\triangle$) [Sn] = 0.2 $\%$, ($\Box$) [Sn] = 0.1 $\%$ and 
($\circ$) [Sn] = 0.05 $\%$. 

\vspace{0.25in}
\noindent
{\bf Figure 3:} Pressure dependence of the hole mobility in several 
samples: ($\bullet$) [N] = 0.5 $\%$, ($\nabla$) [Sn] = 0.5 $\%$,
($\triangle$) [Sn] = 0.2 $\%$, ($\Box$) [Sn] = 0.1 $\%$ and ($\circ$) 
[Sn] = $\mbox{0.05 $\%$.}$ The solid lines are the 
results of our calculations. 
Curves 1 and 2 represent the homopolar phonon and LO polar phonon 
contributions, respectively. Curve 3 represents the ionized impurity 
contribution as calculated for N-doped GaSe. 

\vspace{0.25in}
\noindent
{\bf Figure 4:} Pressure dependence of $E_a$. Curves labeled with N,
Sn$_1$ and Sn$_2$ 
$\mbox{correspond}$ to the  nitrogen and tin levels, 
respectively. The solid lines $\mbox{illustrate}$ the theoretically 
calculated dependence 
of $E_a$ according with the proposed model. The inset show a schematic 
comparison of the Coulomb potential and the proposed impurity 
potential model.

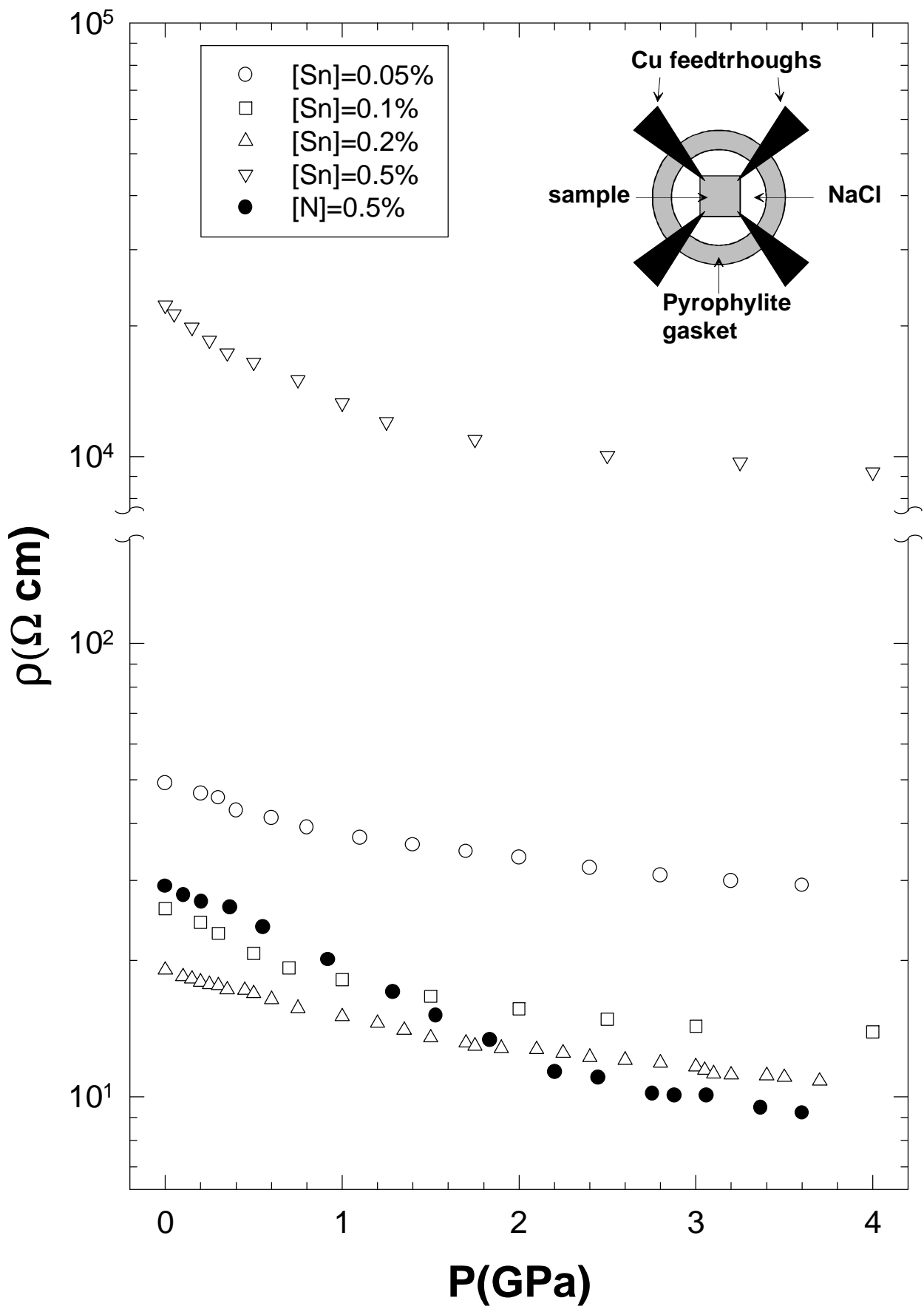
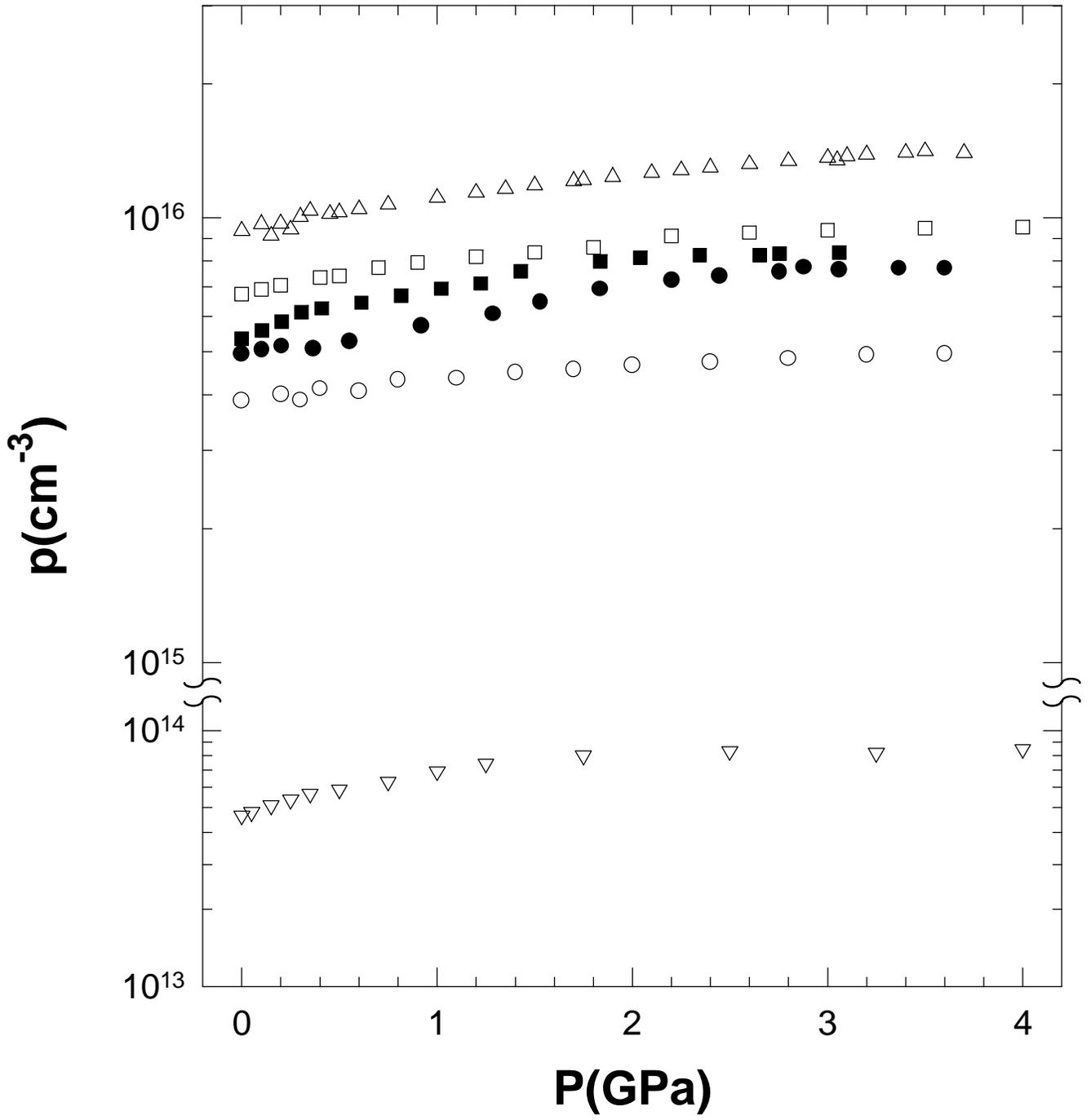
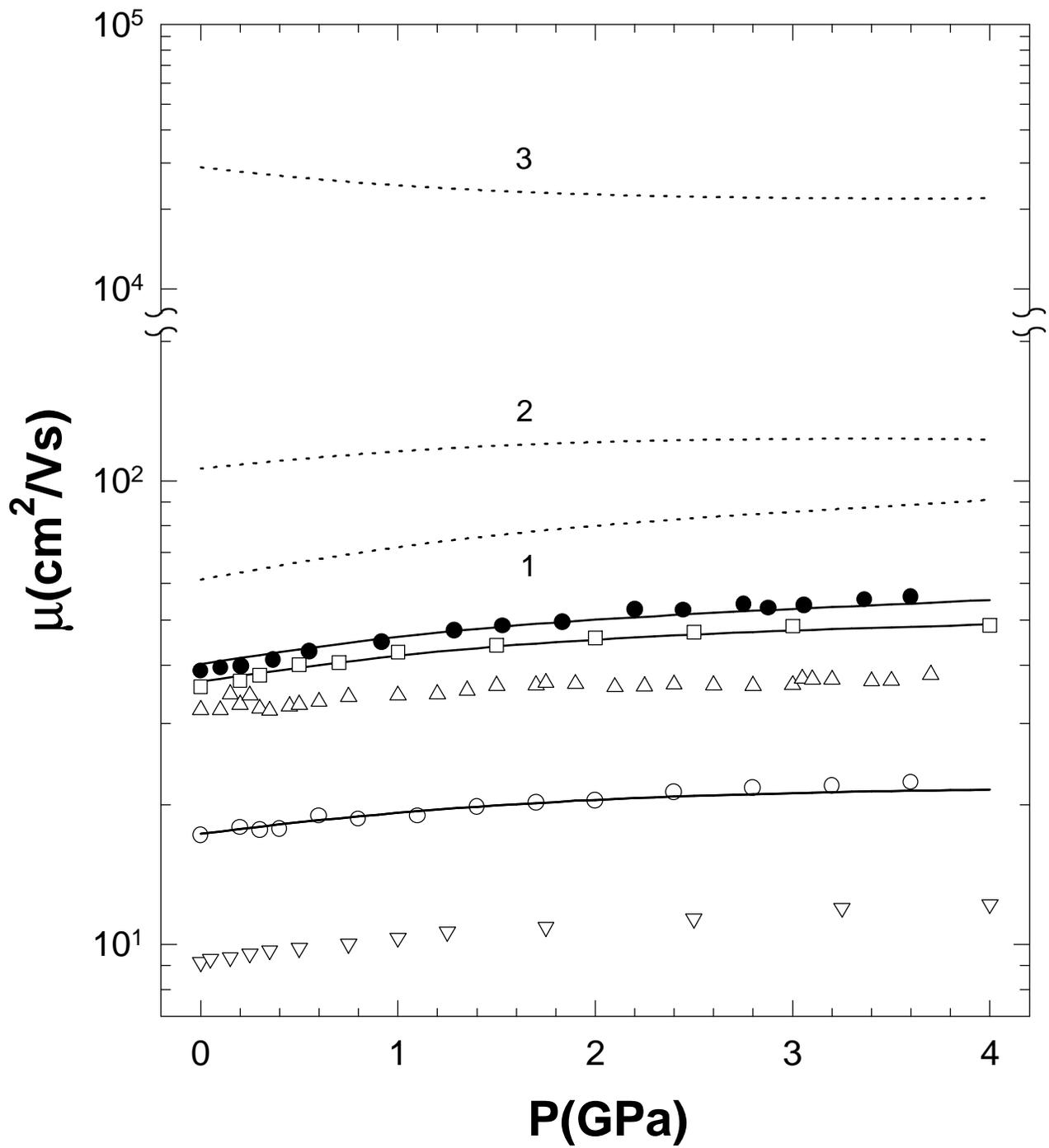
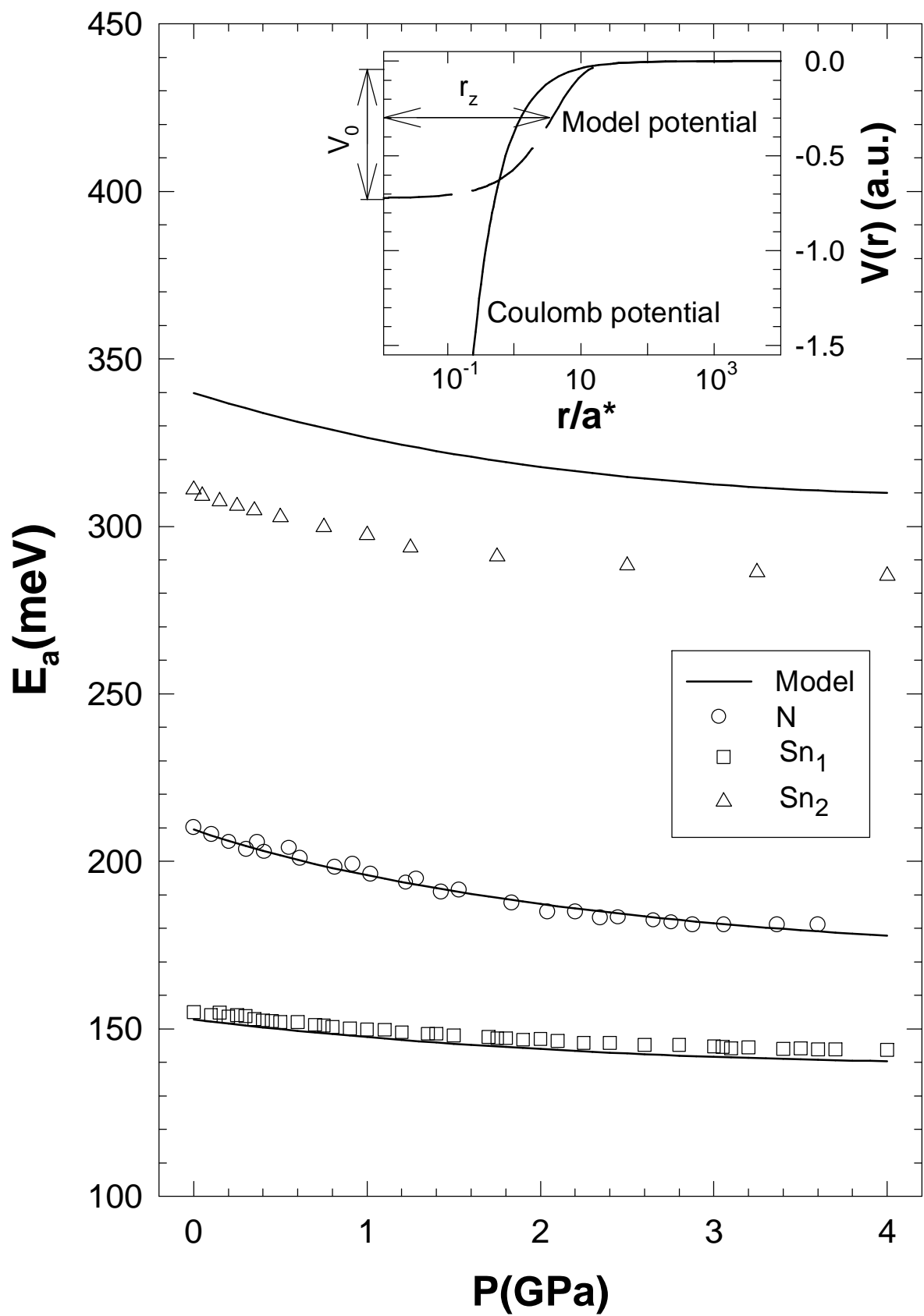
\end{document}